\documentclass[twocolumn,showpacs,preprintnumbers,
amsmath,amssymb,aps,prd,nofootinbib,superscriptaddress,
eqsecnum]{revtex4}
\usepackage{graphicx,color}

\makeatletter
\renewcommand{\p@subsection}{}
\makeatother

\begin{document}

\title{An effective gluon potential and hybrid approach to
Yang-Mills thermodynamics}

\author{Chihiro Sasaki}
\affiliation{%
Frankfurt Institute for Advanced Studies,
D-60438 Frankfurt am Main,
Germany
}

\author{Krzysztof Redlich}
\affiliation{%
Institute of Theoretical Physics, University of Wroclaw,
PL-50204 Wroc\l aw,
Poland
}

\date{\today}

\begin{abstract}
We derive the partition function for the SU(3) Yang-Mills theory
in the presence of a uniform gluon field within  the background
field method. We show, that  the $n$-body gluon contributions in the partition
function are characterized solely by the Polyakov loop.
We express the effective action through characters of  different
representations of the color gauge group resulting in a form deduced in the
strong-coupling expansion. A striking feature of this potential is
that at low temperature gluons are physically disfavored and therefore
they do not yield the correct thermodynamics.
We suggest a hybrid approach to  Yang-Mills  thermodynamics,
combining the  effective gluon potential with
glueballs implemented as dilaton fields.
\end{abstract}


\maketitle

\section{Introduction}
\label{sec:int}

The  $SU(N_c)$ pure gauge theory has a global $Z(N_c)$ symmetry which is
dynamically broken in the high temperature phase. The Polyakov loop,
defined from the temporal gauge field integrated over the Euclidean time,
plays a role of an order parameter of the $Z(N_c)$ global symmetry~\cite{mclerran}.
Effective Polyakov-loop models~\cite{pisarski,pisarski:lect} have been suggested
as a macroscopic approach to the pure SU(3) gauge theory.  The
thermodynamics that comes out from such models  is qualitatively in agreement
with that obtained in  lattice gauge theories~\cite{lat:eos}.
Alternative approaches are based on the quasi-particle picture of
thermal gluons~\cite{peshier}. A natural extension was carried out
where the gluons propagate in a constant
gluon background and the Polyakov loop appears in the partition function
reflecting the group characters~\cite{turko,meisinger:2002,meisinger,kusaka,slqcd}.

In previous   studies the effective potentials for  gluons
were  approximated into the Boltzmann distribution and expanded in
series of the Polyakov loop to finite order. Such expansion  is however unclear
to be justified in the SU(3) gauge theory  around the first-order phase transition.

In this work, our main scope is to derive  the SU(3) gluon thermodynamics utilizing
the background field method for the one-loop quantization and
to clarify the Polyakov-loop effective potentials used in the literature
anchored to the field theoretical basis.
We show,  that {in this approach the   calculated} gluon  potential,  exhibits the correct asymptotic behavior
at high temperatures,
whereas at low temperatures, it disfavors gluons as appropriate dynamical degrees
of freedom.
We  derive  its correspondence to the strong-coupling
expansion, of which the relevant coefficients of the gluon
energy distribution are specified solely by characters of the SU(3) group.

The paper is organized as follows:
In Section~\ref{sec:gluon} we derive the thermodynamic potential for the SU(3)
pure Yang-Mills theory in terms of the Polyakov loop in the fundamental
representation. A correspondence to other phenomenological potentials is
deduced in Section~\ref{sec:asym}.
In Section~\ref{sec:model} we introduce a hybrid approach
that matches  gluons   with   glueballs
at deconfinement transition and study  its thermodynamics.
Our concluding remarks are given in Section~\ref{sec:conc}.

\section{Modeling gluons in hot matter}
\label{sec:gluon}

To formulate thermodynamics
of the  SU(3) Yang-Mills theory,  we start
with the partition function for gluon $A_\mu$ and ghost $C$ fields,
\begin{eqnarray}
Z
&=&
\int{\mathcal D}A_\mu{\mathcal D}C{\mathcal D}\bar{C}\,
\exp\left[i\int d^4x {\mathcal L}\right]\,,
\nonumber\\
{\mathcal L}
&=&
{\mathcal L}_{\rm kin} + {\mathcal L}_{\rm GF}
{}+ {\mathcal L}_{\rm FP}\,,
\end{eqnarray}
with the gauge fixing (GF) and the Faddeev-Popov ghost (FP) terms.
The kinetic term is given by
\begin{eqnarray}
{\mathcal L}_{\rm kin}
&=&
-\frac{1}{2g^2}\mbox{tr}\left[ A_{\mu\nu}A^{\mu\nu}\right]\,,
\nonumber\\
A_{\mu\nu}
&=&
\partial_\mu A_\nu - \partial_\nu A_\mu
{}- i\left[ A_\mu, A_\nu \right]\,,
\end{eqnarray}
where $A_\mu = A_\mu^a T^a$ and
$\mbox{tr}\left[ T^a T^b \right] = \delta^{ab}/2$.

Following~\cite{GPY,pisarski:lect} we employ the background field gauge to
evaluate  the functional integrals. The gluon field is decomposed into background
(classical) $\bar{A}_\mu$ and the quantum  $\check{A}_\mu$  contribution
\begin{equation}
A_\mu = \bar{A}_\mu + g\check{A}_\mu\,.
\end{equation}
We fix the background field gauge with the following form
\begin{equation}
{\mathcal L}_{\rm GF}
= -\frac{1}{\alpha}\mbox{tr}
\left[ \left(\bar{D}_\mu \check{A}^\mu\right)^2 \right]\,,
\end{equation}
where $\alpha$ is the gauge fixing parameter and
$\bar{D}_\mu \check{A}^\nu = \partial_\mu\check{A}^\nu
{}- i\left[ \bar{A}_\mu, \check{A}^\nu\right]$.
Expanding the Lagrangian and collecting the terms quadratic in the quantum
fields one gets
\begin{eqnarray}
{\mathcal L}^{(2)}
&=&
-\frac{1}{2}\check{A}_\alpha^a\left[
\left( D_\mu D^\mu \right)^{\alpha\beta}_{ab}
{}+ \Sigma^{\alpha\beta}_{ab}
\right]\check{A}_\beta^b
\nonumber\\
&&
{}+ i\bar{C}^a \left(D_\mu D^\mu \right)^{(CC)}_{ab}C^b\,,
\label{L2}
\end{eqnarray}
where we define
\begin{eqnarray}
\left(D_\mu\right)^{\alpha\beta}_{ab}
&=&
-g^{\alpha\beta}\delta_{ab}\partial_\mu
{}+ \Gamma_{\mu,ab}^{\alpha\beta}\,,
\nonumber\\
\Gamma_{\mu,ab}^{\alpha\beta}
&=&
-2i\mbox{tr}\left[ \bar{A}_\mu\left[ T_a, T_b\right]\right]
g^{\alpha\beta}\,,
\nonumber\\
\Sigma^{\alpha\beta}_{ab}
&=&
-4i\mbox{tr}\left[ \bar{A}^{\alpha\beta}
\left[ T_a, T_b \right]\right]\,,
\nonumber\\
\left(D_\mu\right)^{(CC)}_{ab}
&=&
\delta_{ab}\partial_\mu + 2i\mbox{tr}\left[
\bar{A}_\mu\left[ T_a, T_b \right]
\right]\,.
\end{eqnarray}
Here 't Hooft-Feynman gauge ($\alpha=1$) was taken~\footnote{
 The partition function must be independent of gauge, i.e.  $d\ln Z/d\alpha=0$.
 Since the running coupling
 depends also on $\alpha$, the condition reads,  $d\ln Z/d\alpha =
 (\partial/\partial\alpha +
 (\partial g/\partial\alpha)(\partial/\partial g)
 )\ln Z = 0$~\cite{kapusta}. In this paper we work in Feynman gauge
 since the standard partition function of a free boson gas is readily
 obtained in the high temperature limit.
}.
Note that the ghost term does not contain $\check{A}_\mu$ and
therefore the Gaussian integral over the ghost fields can easily be
carried out. In the following we keep only the terms quadratic in
$\check{A}_\mu$ and rewrite  Eq.~(\ref{L2})  as
\begin{eqnarray}
{\mathcal L}^{(2)}
&=&
-\frac{1}{2}\check{A}_\alpha^a\left[
\delta_{ab}g^{\alpha\beta}\partial^2
{}- f_{abc}\left( \partial^\beta \bar{A}^{\alpha,c}
{}+ 2g^{\alpha\beta}\bar{A}_\mu^c \partial^\mu \right)
\right.
\nonumber\\
&&
\left.
{}+ f_{ac\bar{c}}f_{cb\bar{d}}g^{\alpha\beta}
\bar{A}_\mu^{\bar{c}} \bar{A}^{\mu,{\bar{d}}}
{}+ 2f_{abc}\bar{A}^{\alpha\beta,c}
\right]\check{A_\beta}^b\,.
\label{L2:2}
\end{eqnarray}
In the above Lagrangian we consider a constant uniform background $\bar{A}_0$:
\begin{equation}
\bar{A}_\mu^a = \bar{A}_0^a\delta_{\mu 0}\,.
\end{equation}
It is convenient to take a diagonal and traceless generators, i.e.
$a=3,8$. We first consider the simplest case where only $\bar A_0^3$ contributes.
Then, in the momentum space,  the Lagrangian  (\ref{L2:2})  is of the following form
\begin{eqnarray}
{\mathcal L}^{(2)}
&=&
-\frac{1}{2}\check{A}_\alpha^b
\left( D^{-1} \right)_{ab}
\check{A}^{\alpha,b}\,,
\nonumber\\
\left(D^{-1}\right)_{ab}
&=&
\delta_{ab}p^2 + 2if_{ab3}\bar{A}_0^3 p_0
{}- f_{ac3}f_{cb3}\left(\bar{A}_0^3\right)^2\,,
\nonumber\\
\end{eqnarray}
where non-vanishing elements of the inverse propagator are given by
\begin{eqnarray}
\left(D^{-1}\right)_{11}
&=&
\left(D^{-1}\right)_{22}
= p^2 + \left(\bar{A}_0^3\right)^2\,,
\nonumber\\
\left(D^{-1}\right)_{33}
&=&
\left(D^{-1}\right)_{88}
= p^2\,,
\nonumber\\
\left(D^{-1}\right)_{44}
&=&
\left(D^{-1}\right)_{55}
= \left(D^{-1}\right)_{66}
= \left(D^{-1}\right)_{77}
\nonumber\\
&=&
p^2 + \frac{1}{4}\left(\bar{A}_0^3\right)^2\,,
\nonumber\\
\left(D^{-1}\right)_{12}
&=&
-\left(D^{-1}\right)_{21}
= 2i\bar{A}_0^3 p_0\,,
\nonumber\\
\left(D^{-1}\right)_{45}
&=&
-\left(D^{-1}\right)_{54}
= -\left(D^{-1}\right)_{67}
= \left(D^{-1}\right)_{76}
\nonumber\\
&=&
i\bar{A}_0^3 p_0\,.
\end{eqnarray}
Diagonalizing  $D^{-1}$ into $\tilde{D}^{-1}=U^\dagger D^{-1} U$ using
unitary transformation,  one finds
\begin{eqnarray}
\label{prop}
\left(\tilde{D}^{-1}\right)_{11}
&=&
\left(p_0 - \bar{A}_0^3\right)^2 - |\vec{p}|^2\,,
\nonumber\\
\left(\tilde{D}^{-1}\right)_{22}
&=&
\left(p_0 + \bar{A}_0^3\right)^2 - |\vec{p}|^2\,,
\nonumber\\
\left(\tilde{D}^{-1}\right)_{33}
&=&
\left(\tilde{D}^{-1}\right)_{88}
= p^2\,,
\nonumber\\
\left(\tilde{D}^{-1}\right)_{44}
&=&
\left(\tilde{D}^{-1}\right)_{77}
= \left(p_0 - \frac{1}{2}\bar{A}_0^3\right)^2 - |\vec{p}|^2\,,
\nonumber\\
\left(\tilde{D}^{-1}\right)_{55}
&=&
\left(\tilde{D}^{-1}\right)_{66}
= \left(p_0 + \frac{1}{2}\bar{A}_0^3\right)^2 -|\vec{p}|^2\,.
\nonumber\\
\end{eqnarray}
The two elements, $(a,b)=(3,8)$, are never mixed with the background
field $\bar{A}_0$ because the generators $T^3$ and $T^8$ are commuting,
$[T^3, T^8] = 0$.

In Euclidean space we replace $p_0$ and $\bar{A}_0$
with $i\omega_n = i2n\pi T$ and $-i\bar{A}_4$, respectively.
Then, with the propagator (\ref{prop}),
the summation over $n$ which appears in the partition function
is  easily performed and  one arrives at~\footnote{
 Equation~(\ref{logdet}) can be generalized to any $N_c$.
 See e.g. \cite{GPY}.
}
\begin{eqnarray}
\label{matsubara}
\sum_n \ln\det\left(D^{-1}\right)
=
\ln\det\left( 1 - \hat{L}_A e^{-|\vec{p}|/T}\right)\,,
\label{logdet}
\end{eqnarray}
with the matrix,
\begin{eqnarray}\label{matrix}
&&
\hat{L}_A
=
\mbox{diag}\left(
1, 1, e^{i\bar{A}^3_4/T}, e^{-i\bar{A}^3_4/T},
\right.
\nonumber\\
&&
\quad\quad
\left.
e^{i\bar{A}^3_4/2T}, e^{-i\bar{A}^3_4/2T},
e^{i\bar{A}^3_4/2T}, e^{-i\bar{A}^3_4/2T}
\right)\,.
\end{eqnarray}
In a  more general case, when
\begin{equation}
\bar{A}_0 = \bar{A}_0^3 T^3 + \bar{A}_0^8 T^8\,,
\end{equation}
the  calculations of  (\ref{matsubara}) can  be extended resulting in the modified
form of the adjoint matrix;
\begin{eqnarray}\label{matrix2}
\hat{L}_A
&=&
\mbox{diag}\left(
1, 1, e^{i\bar{A}^3_4/T}, e^{-i\bar{A}^3_4/T},
\right.
\nonumber\\
&&
\left.
e^{i(\bar{A}^3_4 + \sqrt{3}\bar{A}^8_4)/2T},
e^{-i(\bar{A}^3_4 + \sqrt{3}\bar{A}^8_4)/2T},
\right.
\nonumber\\
&&
\left.
e^{i(\bar{A}^3_4 - \sqrt{3}\bar{A}^8_4)/2T},
e^{-i(\bar{A}^3_4 - \sqrt{3}\bar{A}^8_4)/2T}
\right)\,.
\end{eqnarray}
Since the rank of the SU(3) group  is two,
{the Polyakov loop matrix in the adjoint representation} $\hat{L}_A$, can also be expressed
in terms of two angler parameters $\phi_{1}$ and  $\phi_{2}$ as
\begin{eqnarray}
\hat{L}_A
&=&
\mbox{diag}\left(
1\,, 1\,, e^{i(\phi_1 - \phi_2)}\,,
e^{-i(\phi_1 - \phi_2)}\,,
\right.
\nonumber\\
&&
\left.
e^{i(2\phi_1 + \phi_2)}\,, e^{-i(2\phi_1 + \phi_2)}\,,
e^{i(\phi_1 + 2\phi_2)}\,, e^{-i(\phi_1 + 2\phi_2)}
\right)\,.
\nonumber\\
\end{eqnarray}
Due to this change of variables, the partition function is
rewritten as
\begin{eqnarray}
\ln Z
= V\int\frac{d^3 p}{(2\pi)^3}
\ln\det\left( 1 - \hat{L}_A e^{-|\vec{p}|/T}\right)
{}+ \ln M(\phi_1,\phi_2)\,,
\nonumber\\
\label{parti}
\end{eqnarray}
with $M$ being the Haar measure for a fixed volume $V$ given by
\begin{eqnarray}\label{haar}
M(\phi_1,\phi_2)
&=&
\frac{8}{9\pi^2}\sin^2\left( \frac{\phi_1 - \phi_2}{2}\right)
\sin^2\left( \frac{2\phi_1 + \phi_2}{2}\right)
\nonumber\\
&&
\times
\sin^2\left( \frac{\phi_1 + 2\phi_2}{2}\right)\,,
\end{eqnarray}
which is normalized such that
\begin{equation}
\int_0^{2\pi}\int_0^{2\pi}d\phi_1 d\phi_2 M(\phi_1,\phi_2) = 1\,.
\end{equation}

The first term in Eq.~(\ref{parti})
yields the following thermodynamics potential:
\begin{equation}
\Omega_g
= 2T \int\frac{d^3p}{(2\pi)^3}\mbox{tr}\ln
\left( 1 - \hat{L}_A\, e^{-E_g/T} \right)\,,
\label{omega0}
\end{equation}
where in  the quasi-gluon energy $E_g = \sqrt{|\vec{p}|^2 + M_g^2}$
the  effective gluon mass $M_g$ is introduced from  phenomenological
reasons~\footnote{
  The SU(3) gluon plasma was studied in~\cite{meisinger:2002} starting from
 the same thermodynamic potential as (\ref{omega0}),  whereas the logarithm
 was expanded.
}.
To calculate the thermodynamic potential (\ref{omega0}) one still needs to
perform the trace in a color space.
We define gauge invariant quantities,
normalized by  dimensions of representations, as
\begin{eqnarray}
\Phi = \frac{1}{N_c}\mbox{tr}\hat{L}_F\,,
\quad
\bar{\Phi} = \frac{1}{N_c}\mbox{tr}\hat{L}_F^\dagger\,,
\quad
\Phi_A = \frac{1}{N_c^2-1}\mbox{tr}\hat{L}_A\,,
\nonumber\\
\end{eqnarray}
where $\hat{L}_F$ is the Polyakov loop matrix in the fundamental representation,
\begin{equation}
\hat{L}_F
= \mbox{diag}\left(
e^{i\phi_1}\,, e^{i\phi_2}\,, e^{-i(\phi_1 + \phi_2)}
\right)\,,
\end{equation}
and $\Phi_A$ is related with $\Phi$ and $\bar{\Phi}$ via
\begin{equation}
\left( N_c^2 - 1 \right)\Phi_A = N_c^2 \bar{\Phi}\Phi - 1\,.
\end{equation}

Carrying out the trace over colors and expressing it
in terms of
$\Phi$ and its conjugate $\bar\Phi$,
one finally finds
\begin{equation}
\Omega_g
= 2T \int\frac{d^3p}{(2\pi)^3}\ln
\left( 1 + \sum_{n=1}^8C_n\, e^{-nE_g/T}
\right)\,,\label{gluon}
\end{equation}
with the coefficients $C_n$  given by
\begin{eqnarray}\label{coeff}
C_8
&=&
1\,,
\nonumber\\
C_1
&=&
C_7
= 1 - 9\bar{\Phi}\Phi\,,
\nonumber\\
C_2
&=&
C_6
= 1 - 27\bar{\Phi}\Phi
{}+ 27\left( \bar{\Phi}^3 + \Phi^3\right)\,,
\nonumber\\
C_3
&=&
C_5
= -2 + 27\bar{\Phi}\Phi
{}- 81\left( \bar{\Phi}\Phi \right)^2\,,
\nonumber\\
C_4
&=&
2\left[
-1 + 9\bar{\Phi}\Phi - 27\left( \bar{\Phi}^3 + \Phi^3\right)
{}+ 81\left( \bar{\Phi}\Phi \right)^2
\right]\,.
\nonumber\\
\end{eqnarray}
Thus, the $n$-body gluon contributions to the thermodynamic potential 
(\ref{gluon}) are characterized solely by the
Polyakov loop, i.e. the characters of the fundamental and the conjugate
representations of the color SU(3) gauge group.

The Haar measure (\ref{haar}) is also expressed in terms of $\Phi$ and $\bar{\Phi}$ as
\begin{eqnarray}
&&
M(\phi_1,\phi_2)
\nonumber\\
&&
= \frac{8}{9\pi^2}\left[
1 - 6\bar{\Phi}\Phi + 4\left( \Phi^3 + \bar{\Phi}^3\right)
{}- 3\left(\bar{\Phi}\Phi\right)^2
\right]\,.
\end{eqnarray}

A complete effective thermodynamic potential of gluons in the large volume
limit is obtained from Eq.~(\ref{parti}) as follows:
\begin{eqnarray}\label{full}
\Omega
&=&
\Omega_g + \Omega_\Phi + c_0\,,
\end{eqnarray}
where $\Omega_g $ is given by Eq.~(\ref{gluon}) and
 the Haar measure contribution
\begin{eqnarray}
\Omega_\Phi
&=&
-a_0T\ln\left[ 1 - 6\bar{\Phi}\Phi + 4\left( \Phi^3 + \bar{\Phi}^3\right)
{}- 3\left(\bar{\Phi}\Phi\right)^2\right]\,.
\nonumber\\
\label{dec}
\end{eqnarray}
In (\ref{dec})  we have neglected the normalization factor of the Haar measure
which gives sub-leading contribution to thermodynamics.
The $a_0$ and $c_0$ and/or gluon mass are free parameters  which have to be fixed
through  certain external conditions.
They can be e.g. chosen to reproduce the equation of state of the SU(3)  pure gauge
theory obtained on  the lattice  through  Monte Carlo calculations.
%

\section{Asymptotic  expansions of the potential }
\label{sec:asym}

The  potential  for the SU(3) Yang-Mills theory (\ref{full})
obtained in the previous section  provides  an effective description
of gluon thermodynamics.  In particular, in  asymptotically high temperatures,
the $\Omega_g$  should reproduce the  ideal gas limit.
Indeed, { taking  the limit  $\Phi,\bar{\Phi} \to 1$,  corresponding to $\bar{A}_0 \to 0$,
 one finds  from   (\ref{gluon}), that}
\begin{equation}\label{expansion}
\Omega_g(\Phi=\bar{\Phi}=1)
= 16T \int\frac{d^3p}{(2\pi)^3}
\ln\left( 1 - e^{-E_g/T} \right)\,.
\end{equation}
Thus,  the standard
expression for a non-interacting gas of massive/massless gluons, is recovered.

On the other hand, having in mind a quasi-particle approach, where gluons are
considered as massive particles with a temperature-dependent mass $M_g=M_g(T)$,
one can expand the logarithm in  Eq.~(\ref{gluon}). For a sufficiently large
$M_g(T)/T$ one  approximates the logarithm by the first term of the expansion,
resulting in the following form of the potential;
\begin{equation}
\Omega_g
\simeq {\frac{T^2M_g^2}{\pi^2}}\sum_{n=1}^8{\frac{C_n}{n}}K_2(n\beta M_g)\,,
\label{effective}
\end{equation}
where $C_n$ are as in  Eq.~(\ref{coeff}) and  $K_2(x)$ is the Bessel function.
The above can also be considered as a strong coupling expansion regarding the
relation $M_g(T)=g(T)T$, where
$g(T)$ is  an effective gauge coupling.

The character expansion   of the potential (\ref{effective}) corresponds to
that obtained in  the Polyakov
loop models on the lattice, derived using strong coupling
techniques for  the non-Abelian gauge
group SU(3). Indeed, in the strong-coupling expansion the effective action to
the  next-to-leading
order is obtained in terms of group characters as~\cite{slqcd}~\footnote{
 The action at this order includes one additional term, $\rho_1 V_1$ with
 $V_1 = \sum_{\vec{x}}(|L_F(\vec{x})|^2-1)$~\cite{slqcd}. Since we deal with
 the temporal gluon field as a uniform background, $V_1$ is not distinguished
 from $S_{10}$.
}:
\begin{equation}
S_{\rm eff}^{\rm (SC)}
= \lambda_{10}S_{10} + \lambda_{20}S_{20} + \lambda_{11}S_{11}
{}+ \lambda_{21}S_{21}\,,
\end{equation}
with products of  characters $S_{pq}$, specified by two integers
$p$ and $q$ counting the numbers of fundamental and conjugate
representations, and couplings $\lambda_{pq}$ being real functions of
temperature.
One readily finds those couplings from Eq.~(\ref{effective}), as well as  the
correspondence between $S_{pq}$ and $C_n$ from Eqs.~(\ref{effective})
and (\ref{coeff}), as
\begin{eqnarray}
C_{1,7} = S_{10}\,,
\quad
C_{2,6} = S_{21}\,,
\nonumber\\
C_{3,5} =S_{11}\,,
\quad
C_4 = S_{20}\,.
\end{eqnarray}

Taking only the contribution of a single-gluon distribution $\exp[-M_g/T]$
in the expansion Eq.~(\ref{effective}) yields the ``minimal model'' described by
\begin{equation}
\Omega_g
\simeq -{\mathcal F}(T,M_g)\bar{\Phi}\Phi\,,
\label{approx}
\end{equation}
with the negative sign needed to get a first-order transition as studied in~\cite{slqcd}.
Here, an explicit form of the function ${\mathcal F}$ relies on
approximations used in evaluation of  the momentum integration and
parameterization of   $M_g(T)$. Assuming appropriate temperature dependence
of ${\mathcal F}$ so that the   thermodynamic potential
(\ref{full}) yields
the   phase transition with thermal expectation value of the  Polyakov loop
$\langle\Phi\rangle$ as the order parameter of $Z(3)$ symmetry,
the form widely used in the PNJL model~\cite{pnjl,pnjl:log,pnjl:sus,pnjl:poly} is
recovered.

In addition, the logarithm of the Haar measure part $\Omega_\Phi$  can also be
expanded  in powers of $Z(3)$-invariant operators.
In this case, the effective gluon potential is found in the  polynomial
form~\cite{pisarski,pnjl:poly}. Such form of $\Omega$  can, however,
be applied only to a weak first-order phase transition. The polynomial form
applied in the PNJL  model was also shown to cause some problems in  behaviors
of  charge fluctuations~\cite{pnjl:sus} as well as with the phase structure and
symmetry properties of the potential at complex chemical potential~\cite{morita}.

\begin{figure*}
\begin{center}
\includegraphics[width=8cm]{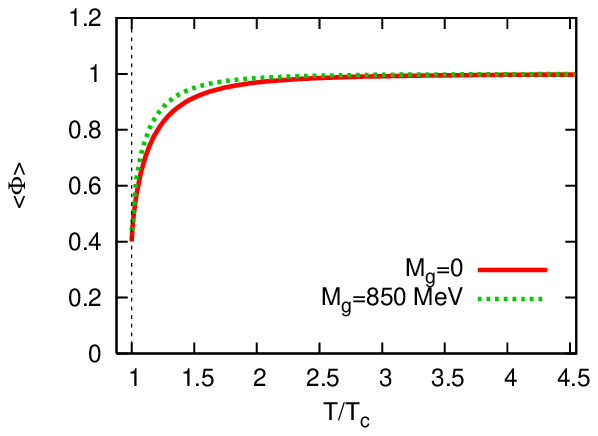}
\includegraphics[width=8cm]{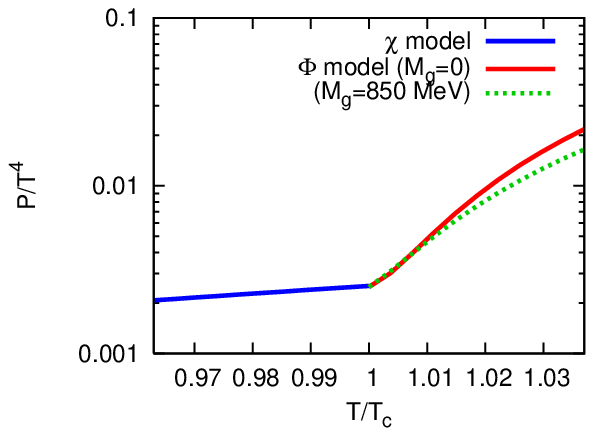}
\caption{
Thermal expectation value of the Polyakov loop (left-hand figure) and
the normalized pressure (right-hand figure)  calculated  in  the hybrid
model (\ref{model}) for massless  $M_g=0$ and massive $M_g=850$ MeV gluons.
}
\label{fig1}
\end{center}
\end{figure*}

\section{A hybrid approach}
\label{sec:model}

The model described by Eqs.~(\ref{gluon}), (\ref{full}) and (\ref{dec})  works
fairly well when the thermal expectation $\langle\Phi\rangle$ of the Polyakov loop
is non-vanishing. However, at any temperatures below $T_c$, where
$\langle\Phi\rangle=0$ is dynamically favored, it causes some unphysical results
on the equation of state.

The model parameters   $a_0$ and $c_0$ are fixed such that the model reproduces
the value of $T_c$ and the pressure at $T_c$ calculated from SU(3) Lattice Gauge Theory.
In such formulation,   the model applied to the phase below $T_c$ yields
 a positive pressure, however the entropy and energy
densities turn out to be negative. In fact, keeping at low temperatures
only the $C_1$ term as a main  contribution, one gets the potential
\begin{equation}
\Omega_g(\Phi=\bar{\Phi}=0)
\simeq 2T \int\frac{d^3p}{(2\pi)^3}
\ln\left( 1 + e^{-E_g/T} \right)\,,
\label{lowT}
\end{equation}
which does not posses   the correct sign in front of $\exp[-E_g/T]$
expected from the Bose-Einstein statistics. One immediately finds  that the
entropy and energy densities calculated from  Eq.~(\ref{lowT}) are negative.
Therefore, the model cannot be naively applied to the phase below $T_c$.
This problem appears not only with the complete potential (\ref{gluon}), but also
with its minimal form (\ref{approx}) which is frequently used in the literature.
There,  the equation of state is entirely zero, at any temperature below $T_c$.
This is clearly unphysical as there are
color-singlet hadrons, glueballs, contributing to thermodynamics in a  pure
Yang-Mills theory and they must generate a non-vanishing pressure.

The above aspects  are in a striking contrast to the quark sector of the thermodynamics
obtained in the presence of a background gluon field $\bar{A}_0$.  There, the
thermodynamic potential for  quarks  and anti-quarks
with  $N_f$ flavors is obtained as,
\begin{eqnarray}
\Omega_{q+\bar q}
&=&
-2N_f T \int\frac{d^3p}{(2\pi)^3}\mbox{tr}\ln
\left[ 1 + \hat{L}_F\, e^{-(E_q-\mu)/T} \right]
\nonumber\\
&&
{}+ \left(\mu\to -\mu\right)\,.
\end{eqnarray}
The trace over color indices in this case is easily performed and
the potential is  expressed by characters of fundamental  $\Phi$ and the conjugate
$\bar{\Phi}$ representation as~\cite{pnjl,megias}
\begin{eqnarray}
\Omega_{q+\bar q}
&=&
-2N_f T \int\frac{d^3p}{(2\pi)^3}
\nonumber\\
&\times&
\ln
\left[ 1 + N_c\left( \Phi + \bar{\Phi}e^{-E^+/T}\right)e^{-E^+/T}
{}+ e^{-3E^+/T}\right]
\nonumber\\
&&
{}+
\left(\mu \to -\mu\right) \,,
\nonumber\\
\label{omega:quark}
\end{eqnarray}
with $E^\pm = E_q \mp \mu$ being  the energy of a quark (anti-quark).

In the limit of $\Phi,\bar{\Phi}\to 0$, which is expected at low
temperatures, the contribution of one- and two-quark states is
suppressed and only  three-quark (baryonic) states,  $\sim\exp(-3E^{(\pm)}/T)$,
survives. This, on a qualitative level, is similar to confinement in QCD
thermodynamics~\cite{pnjl:sus}. One should however keep in mind that
this model  yields only  colored quarks being  {\it statistically} suppressed
at low temperatures.
On the other hand, unphysical thermodynamics that comes out  below $T_c$ from
the gluon sector (\ref{full}), apparently indicates that  gluons
are {\it physically} forbidden bellow $T_c$. 
We note that in the mean field approximation higher representations of 
the Polyakov loop than the fundamental one are non-vanishing below $T_c$. 
This is an artifact 
of the approximation and in fact they do not condense when one evaluates 
their group averages with the Haar measure. Such ``hidden'' physics of the 
higher representations can be embedded in the mean field approach when all 
the gluon energy distributions are expressed in terms of the fundamental 
Polyakov loop $\Phi$. In this way all the colored gluons are suppressed
and therefore the correct physics interpretation is recovered.

This property is not affected by quarks.
Let us consider massive gluons and quarks at zero chemical potential.
From Eqs.~(\ref{lowT}) and (\ref{omega:quark})
applied to $T<T_c$, the thermodynamic potential is approximated as
\begin{equation}
\Omega_g + \Omega_{q+\bar{q}}
\simeq
\frac{T^2}{\pi^2}\left[
M_g^2 K_2\left(\frac{M_g}{T}\right)
{}- \frac{2N_f}{3}K_2\left(\frac{3M_q}{T}\right)
\right]\,.
\label{lowTapprox}
\end{equation}
Assuming that the glueball and nucleon are made from two gluons and
three quarks respectively and putting empirical numbers
$M_{\rm glueball}=1.7$ GeV~\cite{sexton} and $M_{\rm nucleon}=0.94$ GeV,
one finds $M_g = 0.85$ GeV and $M_q = 0.31$ GeV.
Given those numbers, Eq.~(\ref{lowTapprox}) is
positive for either $N_f=2$ or $3$. Consequently, the entropy density
is negative at any temperature as found in a pure Yang-Mills case.
Therefore, thermodynamics remains unphysical, unless additional terms
responsible for the non-perturbative effects in confined phase are
considered.

\subsection{Modeling glueballs as  dilaton fields}

Below $T_c$,  thermodynamics needs to be described in terms of
physical degrees of freedom, i.e. glueballs.
We introduce a glueball as a dilaton field $\chi$ representing the gluon
composite $\langle A_{\mu\nu}A^{\mu\nu}\rangle$,  which is responsible
for the QCD trace anomaly~\cite{schechter}. The Lagrangian that we use is of
a standard form given by
\begin{eqnarray}\label{dilaton}
{\mathcal L}_\chi
&=&
\frac{1}{2}\partial_\mu\chi\partial^\mu\chi - V_\chi\,,
\nonumber\\
V_\chi
&=&
\frac{B}{4}\left(\frac{\chi}{\chi_0}\right)^4
\left[ \ln\left(\frac{\chi}{\chi_0}\right)^4 - 1 \right]\,,
\end{eqnarray}
where $B$ is the bag constant and $\chi_0$ is a dimensionful quantity.
The two parameters $B$ and $\chi_0$ can be fixed
using the vacuum energy density ${\mathcal E} = \frac{1}{4}B = 0.6$
GeV fm$^{-3}$~\cite{narison} and the vacuum glueball mass $M_\chi = 1.7$
GeV~\cite{sexton} with the following definition:
\begin{equation}
M_\chi^2 = \frac{\partial^2V_\chi}{\partial\chi^2}\Big{|}_{\chi=\chi_0}
= \frac{4B}{\chi_0^2}\,.
\end{equation}
One finds that
$B = (0.368\,\mbox{GeV})^4$ and
$\chi_0 = 0.16\,\mbox{GeV}$.

With the Lagrangian (\ref{dilaton}),  the thermodynamic potential of
effective glueball fields is found to be
\begin{eqnarray}
\Omega
&=&
\Omega_\chi + V_\chi + \frac{B}{4}\,,
\nonumber\\
\Omega_\chi
&=&
T\int\frac{d^3p}{(2\pi)^3}\ln\left(1-e^{-E_\chi/T}\right)\,,
\nonumber\\
E_\chi
&=&
\sqrt{|\vec{p}|^2 + M_\chi^2}\,,
\quad
M_\chi^2 = \frac{\partial^2 V_\chi}{\partial\chi^2}\,,
\label{conf}
\end{eqnarray}
where a constant $B/4$ is added so that $\Omega = 0$ at zero temperature.

\begin{figure*}
\begin{center}
\includegraphics[width=8cm]{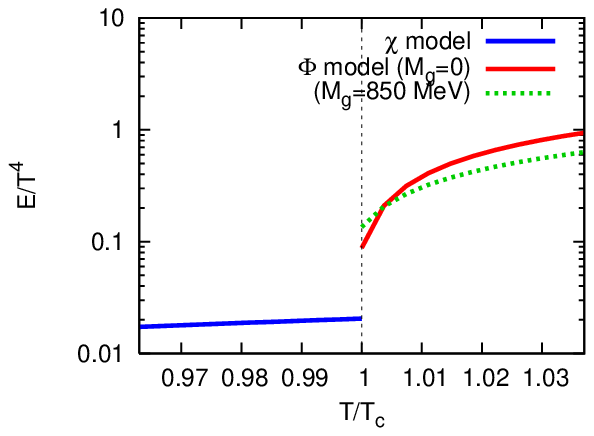}
\includegraphics[width=8cm]{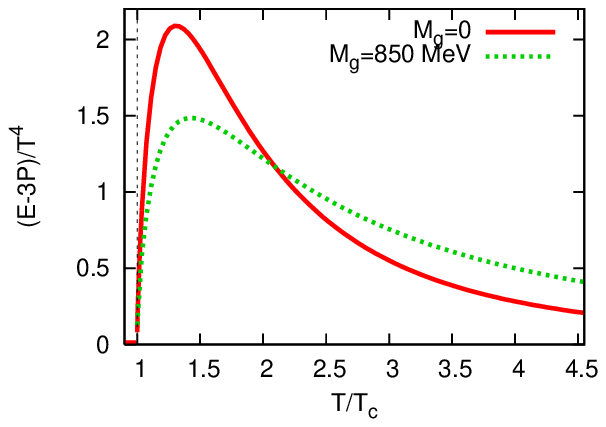}
\caption{
Normalized energy  density (left-hand figure) and the interaction measure
(right-hand figure) calculated  in  the hybrid model (\ref{model}) for massless
$M_g=0$ and massive $M_g=850$ MeV gluons.
}
\label{fig2}
\end{center}
\end{figure*}

\subsection{The hybrid thermodynamic potential}

To avoid problems of unphysical equations of state in confined phase
we adopt a hybrid approach which accounts  for gluons and glueballs degrees
of freedom by combining Eqs.~(\ref{full}) and
(\ref{conf}) as follows:
\begin{equation}
\Omega = \Theta(T_c-T)\,\Omega(\chi) + \Theta(T-T_c)\,\Omega(\Phi)\,.
\label{model}
\end{equation}
The model parameters are constrained by requiring that
\begin{itemize}
\item
$\Omega(\Phi)$ yields a first-order phase transition at $T_c=270$ MeV
as found in SU(3) lattice calculations~\cite{lat:eos,lat:poly}.
\item
$\Omega(\chi)$ and $\Omega(\Phi)$ match at $T_c$.
\end{itemize}
When the gluon effective mass is assumed to be zero, one finds  the following
model parameters:
\begin{eqnarray}
M_g=0:
&&
\langle\Phi\rangle_{T_c} = 0.395\,,
\quad
a_0 = (0.197\,\mbox{GeV})^3\,,
\nonumber\\
&&
c_0 = -(0.180\,\mbox{GeV})^4\,.
\end{eqnarray}
One can also assume that the gluon becomes massive via non-vanishing gluon
condensation $\langle\chi\rangle \neq 0$~\cite{carter}. Requiring that
a glueball is composed of two constituent gluons yields
$M_g = M_\chi/2$. Since $\langle\chi\rangle$ little varies with
temperature around $T_c$~\cite{cdm}, we treat $M_g$ as a constant and
choose $M_g = 1.7\,\mbox{GeV}/2 = 0.85$ GeV.
The parameters in this case are found as
\begin{eqnarray}
M_g=0.85\,\mbox{GeV}:
&&
\langle\Phi\rangle_{T_c} = 0.439\,,
\quad
a_0 = (0.125\,\mbox{GeV})^3\,,
\nonumber\\
&&
c_0 = -(0.130\,\mbox{GeV})^4\,.
\end{eqnarray}
A more general case, not considered in this paper,  would include the
temperature-dependent effective gluon mass which is fixed such that
the present model quantifies
thermodynamics calculated on the lattice in the  SU(3) gauge theory.

\subsection{Thermodynamics}
\label{sec:eos}

Thermodynamic properties of the hybrid model and its phase structure can be
quantified  directly from the potential (\ref{model}).
Fig.~\ref{fig1} shows  the thermal expectation value  of the
Polyakov loop $\langle\Phi\rangle$  obtained from Eq.~(\ref{model}) as the
solution of the stationary condition, $\partial \Omega/\partial \Phi=0$.
There is a trivial solution $ \langle\Phi\rangle=0$ at any temperatures and
it becomes degenerate with a non-trivial solution $\langle\Phi\rangle$
at some $T_c$, indicating a first-order deconfinement transition.

The Polyakov loop expectation   is weakly  changing  with $M_g$
and approaches unity rather quickly as seen  in Fig.~\ref{fig1}.
The temperature dependence of $\langle\Phi\rangle$ just above $T_c$  and its
value at $T_c$ are  consistent with lattice results and can be still improved by
introducing a thermal gluon mass $M_g(T)$ as done e.g. in \cite{meisinger}.
There, the effective mass was parameterized
as in the standard quasi-particle approaches at high temperature,
$M_g(T) = g(T)T$, with the  effective  running coupling $g(T)$~\cite{peshier}.
The lattice data on the renormalized $\langle\Phi\rangle$ are known to exceed
unity at $T/T_c \sim 3$ \cite{kaczmarek}. This property of lattice data, which is
associated with uncertainties of the renormalization
procedure~\cite{pqcd,pisarski:lect}, can never appear in effective
Polyakov loop  models where $\Phi$ is  the character of the fundamental
representation and restricts the target space, so that $\langle\Phi\rangle$
is not allowed to go beyond unity.

Fig.~\ref{fig1}-right   shows the pressure calculated from the effective gluon
(\ref{full}) and from  the effective glueball
(\ref{conf}) potential for massless and massive gluons.
Although the presence of a constant $c_0$ in Eq.~(\ref{full}) makes the
pressure  positive below $T_c$,
as mentioned in the previous section, Eq.~(\ref{full}) unavoidably
leads to a negative entropy and  energy densities.  Consequently, in  the
hybrid model (\ref{model}) and for $T\leq T_c$,  the pressure must be
quantified by the  glueball potential (\ref{conf}).
The cusp at $T_c$ in pressure implies a discontinuity in its temperature derivative.
The energy density and the interaction measure $\Delta=(E -3P)/T^4$ are presented
in Fig.~\ref{fig2}.
The energy density has a jump from glueballs to gluons thermodynamics at $T_c$,
whereas the interaction measure exhibits a maximum just above $T_c$.

Even though, the qualitative behaviors of $E/T^4$  and $\Delta$  follow general
trends seen in  lattice data,
the EoS is apparently more sensitive to $M_g$ than $\langle\Phi\rangle$.
The model calculations  with  massless gluons converge too  quickly to asymptotic
Stefan-Boltzmann  limit. The EoS for massive gluons,  on the other hand,
exhibits  a better agreement with  lattice data.
This clearly indicates the presence of some residual
interactions above  $T_c$ which can be {incorporated}  into $M_g(T)$. Thus, to quantify
lattice results one would need to include the temperature-dependent gluon mass.
In simplified quasi-particle models,  where  $\Phi=1$ for any $T$, the $M_g(T)$
was shown to be strongly increasing when approaching $T_c$ from above.
This behavior, however, can be modified if the contribution of the background
gauge field   is included in the quasiparticle model formulated in Eq.~(\ref{gluon}).

For  $T\leq T_c$ the hybrid model includes a glueball as the relevant
degree of freedom, similarly as seen in  lattice calculations. However,
in the present treatment, the model contains only the lowest-lying glueball,
which  might not be sufficient to quantify lattice thermodynamics in a confined
phase. One way out  would be to deal with a gluon degeneracy factor as an additional
parameter.

\section{Conclusions}
\label{sec:conc}

We have derived the thermodynamic potential in the  SU(3) pure Yang-Mills
theory in the presence of a uniform gluon field within  the background
field method. We have shown  that such effective gluon   potential, which
accounts for quantum statistics and reproduces an  ideal gas limit at high
temperatures, is formulated
in terms of the Polyakov loop in the fundamental representation.

The gluon distributions are found to be specified solely by the Polyakov
loop and therefore  there is  one-to-one correspondence to the effective
action in the strong-coupling expansion.
We have shown that  effective models of the Polyakov loop used so far to
describe pure gauge theory thermodynamics appear  as  limiting cases of our result.

Our main observation is that  the effective Polyakov-loop   potential can not be
applied to the phase where the thermal expectation value of the Polyakov-loop
vanishes.
There, in confined phase, the equation of state is unphysical resulting in negative
entropy and energy densities.
This property of gluon potential
is in remarkable contrast to the description of ``confinement''
within a class of chiral models with Polyakov loops~\cite{pnjl,pnjl:poly}. There,
colored quarks are  suppressed only {\it statistically} at low
temperatures.

The gluonic model considered here
indicates that colored gluons are forbidden below $T_c$ as dynamical
degrees of freedom. This feature  is unchanged by the presence of quarks.
To avoid problems of unphysical thermodynamics in confined phase
we proposed a hybrid approach which matches at deconfinement critical
temperature    the effective model of  gluons to the one of glueballs
constrained by the QCD trace anomaly.

The approach developed in this work is open to further investigations of
the SU(3) gluodynamics guided by available lattice results,  as well as  to
a more realistic description of an effective  QCD thermodynamics with quarks.

\subsection*{Acknowledgments}

We acknowledge stimulating discussions with Bengt Friman.
The work of C.S. has been partly supported by the Hessian LOEWE initiative
through the Helmholtz International Center for FAIR (HIC for FAIR).
K.R. acknowledges support by the Polish Science Foundation (NCN).


\end{document}